\documentclass[slac_one]{revtex4}
\usepackage{graphicx}
\usepackage{fancyhdr}
\begin{document}

\title{A new kind of weak-coupling in top-quark physics?} 

%

\author{Charles A. Nelson} \affiliation{Department of Physics, State University of New
York at Binghamton, Binghamton, N.Y. 13902}

\begin{abstract}
In the standard model (SM), for the $ t \rightarrow  W+ b $ decay
mode, the relative phase is $ 0^o $ between the dominant $
A(0,-1/2) $ and $ A(-1, -1/2) $ helicity amplitudes. However, in
the case of an additional large $ t_R \rightarrow b_L $ chiral
weak-transition moment, there is instead a $ 180^o $ relative
phase and three theoretical numerical puzzles. This phase can be
measured at the Tevatron or LHC in top-antitop pair production by
use of W-boson longitudinal-transverse interference in
beam-referenced stage-two spin-correlation functions. Indeed, this
is a new type of weak-coupling for it is directly associated with
$ E_W $, the W-boson energy in the top quark rest frame, instead
of with a canonical effective mass scale. For most $ 2 \rightarrow
2 $ reactions, the simple off-shell continuation of this
additional coupling is found to have good high energy properties,
i.e. it does not destroy 1-loop unitarity of the SM. In a subset
of processes, additional third-generation couplings are required.
\end{abstract}

\maketitle

\thispagestyle{fancy}


\section{Does the top-quark have a large chiral weak-transition
moment in $t\rightarrow W^+ b $ decay?}

By W-boson longitudinal-transverse quantum interference, the
relatively simple 4-angle beam-referenced stage-two
spin-correlation function
\begin{equation}
{{\mathcal{G}}{|}}_{0} +  {{\mathcal{G}}{|}}_{sig}
\end{equation}
enables measurement of the relative phase of the 2 dominant
amplitudes in $ t\rightarrow W^{+}b $ decay with both gluon and
quark production contributions [1-4].

In the standard model, for the $t \rightarrow W^+ b$ decay mode,
the relative phase is $0^o$ between the dominant $A(0,-1/2)$ and
$A(-1,-1/2)$ helicity amplitudes for the standard model $V-A$
coupling.

For the case of an additional chiral-tensorial-coupling in $g_L =
g_{f_M + f_E} = 1$ units,
\begin{equation}
\frac{1}{2} \Gamma ^\mu =g_L \left[ \gamma ^\mu P_L + \frac{1 }
{2\Lambda _{+} }\imath \sigma ^{\mu \nu } ({q_W})_\nu P_R \right]
\nonumber \\
= P_R \left[ \gamma ^\mu + \iota \sigma ^{\mu \nu } v_\nu \right]
\end{equation}
where  $P_{L,R} = \frac{1}{2} ( 1 \mp \gamma_5 ) $ and $\Lambda_+
= E_W / 2 \sim 53 GeV$ in the top rest frame. In the case of such
an additional large $t_R \rightarrow b_L$ chiral weak-transition
moment, there is instead a $180^o$ relative phase between the
$A(0,-1/2)$ and $A(-1,-1/2)$ helicity amplitudes. The associated
on-shell partial-decay-width $\Gamma( t \rightarrow W^+ b )$ does
differ for these two Lorentz-invariant couplings [
$\Gamma_{SM}=1.55GeV$, $\Gamma_+ = 0.66GeV$ versus less than $12.7
GeV @ 95\% C.L. $ in CDF conf. note 8953(2007) in PDG2008 ].

\section{Helicity decay amplitudes}

In the  $t_1$ rest frame, the matrix element for $t_1 \rightarrow
{W_1}^{+} b$ is
\begin{equation}
\langle \theta _1^t ,\phi _1 ,\lambda _{W^{+} } ,\lambda _b |\frac
12,\lambda _1\rangle =D_{\lambda _1,\mu }^{(1/2)*}(\phi _1 ,\theta
_1^t ,0)A\left( \lambda _{W^{+} } ,\lambda _b \right)
\end{equation}
where $\mu =\lambda _{W^{+} } -\lambda _b $ in terms of the
${W_1}^+$ and $b$-quark helicities. An asterisk denotes complex
conjugation.  The final ${W_1}^{+}$ momentum is in the $\theta
_1^t ,\phi _1$ direction and the $b$-quark momentum is in the
opposite direction.  We use the Jacob-Wick phase convention.

So there are 4 moduli and 3 relative phases to be measured, see
refs. in [1]. Both in the SM and in the case of an additional
large $t_R \rightarrow b_L $ chiral weak-transition moment, the
$\lambda _{b}=-1/2$ and $ \lambda \overline{_{b}}$ $=1/2$
amplitudes are more than $\sim 30$ times larger than the $\lambda
_{b}=1/2$ and $\lambda \overline{_{b}}$ $=-1/2$ amplitudes.
\begin{table}[hbt]\centering
\vskip -2.cm
\includegraphics[width=12.cm]{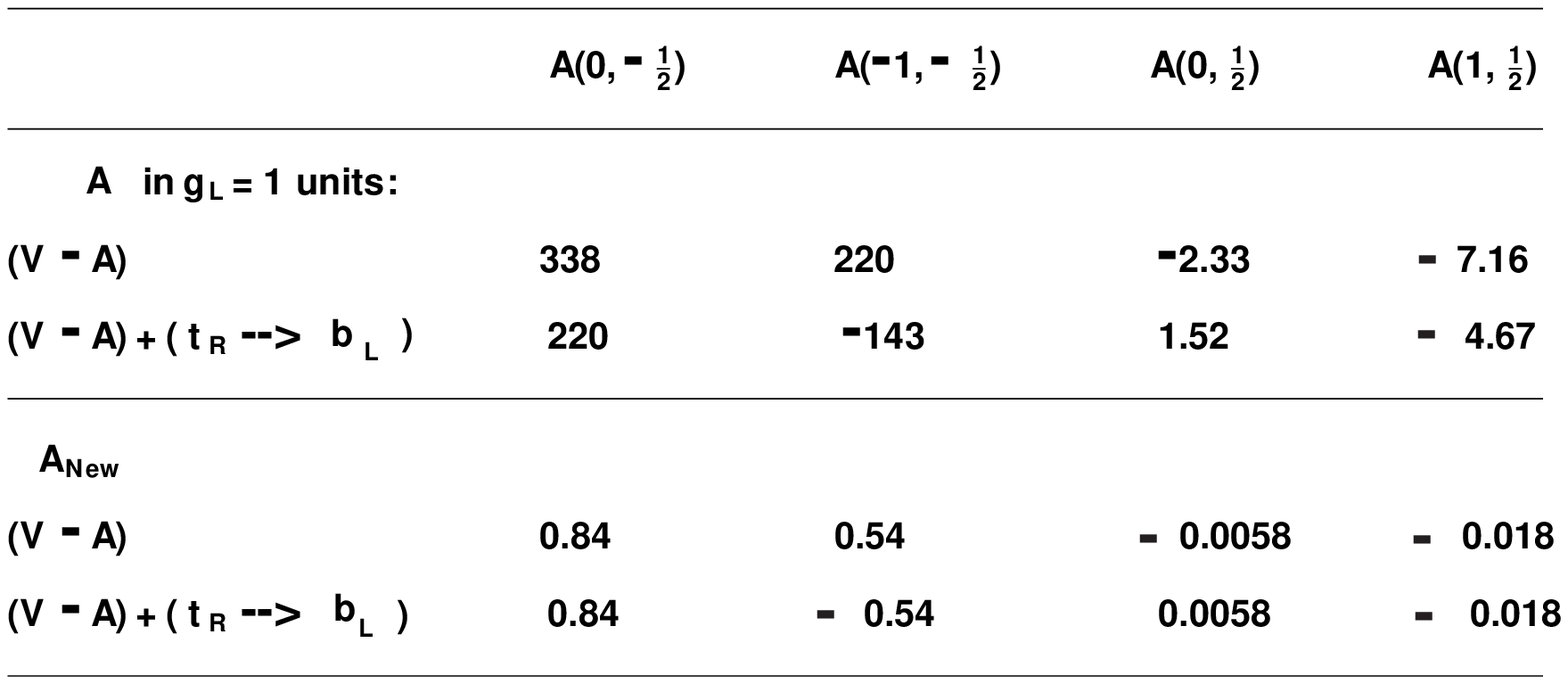}
\caption{Helicity amplitudes for $(V-A)$ coupling and the $(+)$
coupling of Eq.(2). Note $A_{New} = A_{g_L =1} / \sqrt \Gamma $.}
\end{table}

\section{ Idea of a W-boson Longitudinal-Transverse interference measurement }

For the charged-lepton-plus-jets reaction $p p $ or $p \bar{p}
\rightarrow t \bar{t} \rightarrow (W^+ b) (W^- \bar{b} )
\rightarrow  (l^{+} \nu b ) (W^- \bar{b} )$, one selects a
``signal contribution" so that its intensity-observable is the
product of an amplitude in which the $W^+$ is
longitudinally-polarized with the complex-conjugate of an
amplitude in which the $W^+$ is transversely polarized, summed
with the complex-conjugate of this product. The helicity formalism
is a general method for investigating applications of W-boson
interference in stage-two spin-correlation functions for
describing the charged-lepton plus jets channel, and for the
di-lepton plus jets channel.

\section{ Observables and signatures }

The 2 dominant polarized partial widths are
\begin{eqnarray}
\Gamma (0,0) & \equiv &\left| A(0,-1/2)\right| ^{2}, \, \, \Gamma
(-1,-1)\equiv \left| A(-1,-1/2)\right| ^{2}
\end{eqnarray}

The 2 W-boson Longitudinal-Transverse interference widths are
\begin{eqnarray}
\Gamma _{\mathit{R}}(0,-1) &=&\Gamma _{\mathit{R}}(-1,0)\equiv {{
Re}[A(0,-1/2)A(-1,-1/2)^{\ast }}]  \nonumber \\
& \equiv & |A(0,-1/2)||A(-1,-1/2)|\cos \beta _{L}  \\
\Gamma _{\mathit{I}}(0,-1) &=&-\Gamma _{\mathit{I}}(-1,0) \equiv {Im}%
[A(0,-1/2)A(-1,-1/2)^{\ast }]  \nonumber \\
& \equiv &-|A(0,-1/2)||A(-1,-1/2)|\sin \beta _{L}
\end{eqnarray}
The relative phase of these 2 dominant amplitudes is $\beta _{L}$.
In both models
\begin{equation}
Probability \, \, W_{L}=\frac{\Gamma (0,0)}{\Gamma }=0.70
\end{equation}
\begin{equation}
Probability \, \, W_{T}=\frac{\Gamma (-1,-1)}{\Gamma }=0.30
\end{equation}

But there are the respective signatures
\begin{eqnarray}
\eta _{L}  \equiv  \frac{ \Gamma_R (0,-1) } { \Gamma } = + \, 0.46
\, (SM)  \nonumber
\end{eqnarray}
 for the Standard Model, and
\begin{eqnarray}
\eta _{L}   = - \, 0.46 \, (+) \nonumber
\end{eqnarray}
for the case of an additional large chiral weak-transition moment.
In both models, unless there is a violation of time-reversal
invariance
\begin{equation}
\eta _{L}^{^{\prime }}\equiv \frac{\Gamma _{I}(0,-1)}{\Gamma }=0
\end{equation}

\section{ Definition of angles in spin-correlation function }

For $p p $ or $p \bar{p} \rightarrow t \bar{t} \rightarrow (W^+ b)
(W^- \bar{b} ) \rightarrow  (l^{+} \nu b ) (W^- \bar{b} )$, the
``Spin-Correlation Function'' depends on four angles. While the
cosine of the \textbf{4th angle}, $\cos \Theta _{B}$ , can be
integrated out, the expressions are clearer if it is not.  $\Theta
_{B}$ is the ``beam referencing angle" in the $(t\
\overline{t})_{cm}$ frame [3,4].

\textbf{The 3 angles are:}

(i) The spherical angles $\mathbf{\theta }_{a}$ and $\phi _{a}$
which specify the final positive-charged lepton in the $W_{1}^{+}$
rest frame when the boosts are from the $(t\ \overline{t})_{cm}$
frame to the $t_1$ frame and then to the $W_{1}^{+}$ rest frame.
\begin{figure}[hbt]\centering
\includegraphics[width=10.cm]{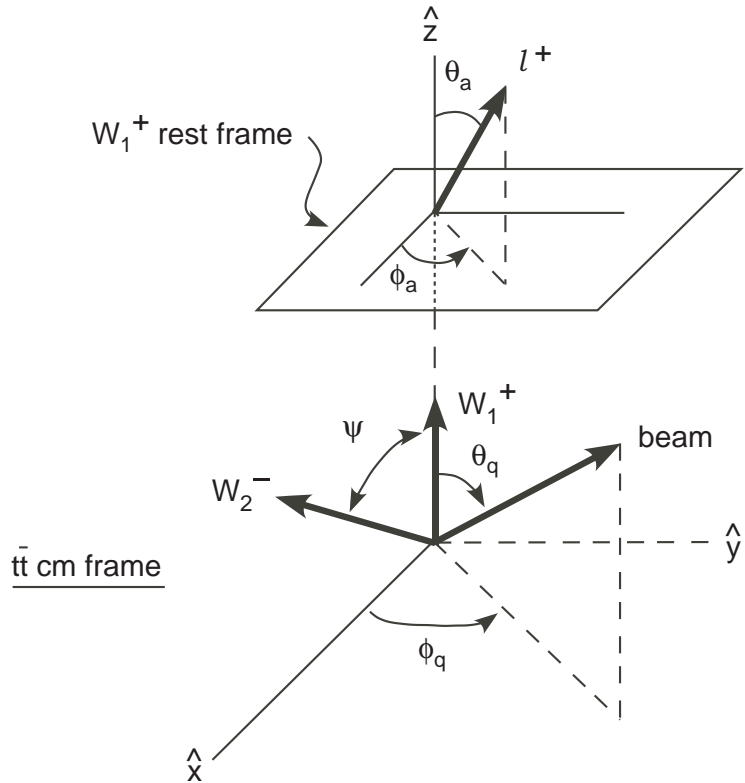}
\caption{The spherical angles $ \theta_a$, $\phi_a $ specify the $
l^+ $ momentum in the ${W_1}^+$ rest frame.}
\end{figure}
\begin{figure}[hbt]\centering
\includegraphics[width=14.cm]{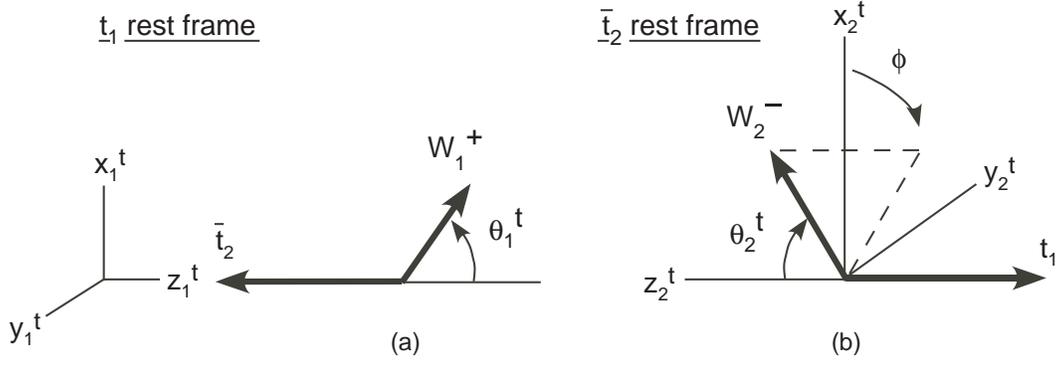}
\caption{Summary illustration showing the three angles $\theta
_1^t$, $\theta _2^t$ and $\phi $ describing the first stage in the
sequential-decays of the $t\bar{t}$ system in which $%
t_1 \rightarrow {W_1}^{+}b$ and $\bar{t_2} \rightarrow
{W_2}^{-}\bar b$. In (a) the $b$ momentum, not shown, is back to
back with the ${W_1}^{+}$.  In (b) the $\bar{b}$ momentum, not
shown, is back to back with the ${W_2}^{-}$.}
\end{figure}
(ii) The cosine of the polar angle $\theta _{2}^{t}$ to specify
the $W_{2}^{-}$ momentum direction in the anti-top rest frame.
Usage of $\cos \theta _{2}^{t}$ is equivalent to using the $(t\
\overline{t})_{cm}$ energy of this hadronically decaying
$W_{2}^{-}$ .
\begin{figure}[hbt]\centering
\includegraphics[width=10.cm]{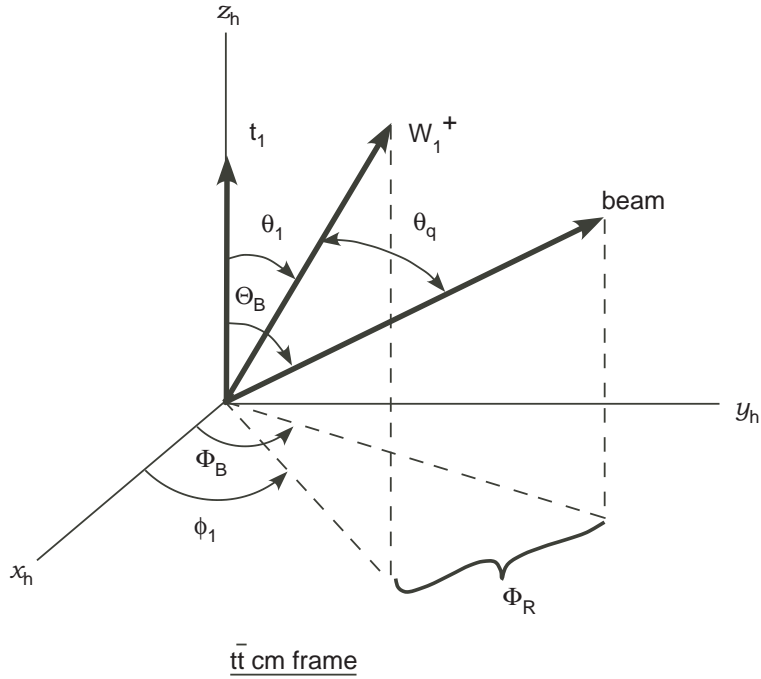}
\caption{The $g_1$ gluon-momentum or $q_1$ quark-momentum ``beam"
direction is specified by the spherical angles $\Theta_B, \Phi_B$.
$\cos \Theta _{B}$ can be integrated out.}
\end{figure}

\section{ Spin-correlation function }

By W-boson longitudinal-transverse interference, the relatively
simple 4-angle beam-referenced stage-two spin-correlation function
\begin{equation}
{{\mathcal{G}}{|}}_{0} +  {{\mathcal{G}}{|}}_{sig}
\end{equation}
enables measurement of the relative phase of the 2 dominant
amplitudes in $ t\rightarrow W^{+}b $ decay:

The ``background term" is
\begin{eqnarray}
{ \mathcal{G}}{|} _{0} &=&  \,\, B(s,\Theta _{B}){|} _{0} \,\,
\left\{ \frac{1}{2}\Gamma (0,0)\sin ^{2}\theta _{a}+\Gamma (-1,-1)\sin ^{4}%
\frac{\theta _{a}}{2}\right\}
\end{eqnarray}

The ``signal term" is
\begin{eqnarray}
{\mathcal{G}}{|}_{sig} &=&{-}\frac{\pi}{2\sqrt{2}} \,\,B(s,\Theta
_{B}){|} _{sig} \,\,  \cos \theta _{2}^{t}\sin \theta _{a}\sin
^{2}\frac{\theta _{a}}{2}
\nonumber \\
&&\left\{ \Gamma _{R}(0,-1)\cos \phi _{a}-\Gamma _{I}(0,-1)\sin
\phi _{a}\right\} {\mathcal{R}}
\end{eqnarray}
The signal contribution is suppressed by
 by the factor ${\mathcal{R}} = ({\mathtt{prob}}
\, W_L) - ({\mathtt{prob}} \, W_T) = 0.40 $. Eqs (11,12) omit a
common overall factor $ \frac{16\pi
^{3}g^{4}}{9 s^{2}}  [\overline{\Gamma }(0,0)+\overline{\Gamma }%
(1,1)] $, see [3,4].

\bigskip

\textbf{Beam-Referencing Factors: }

In Eq.(10), one adds the quark and the gluon production
contributions with for quark-production
\begin{eqnarray}
{ {B}_{0}^{q}(s,\Theta _{B}) } &=& { \frac{1}{24}[1+\cos ^{2}\Theta _{B}+\frac{4m^{2}}{%
s}\sin ^{2}\Theta _{B}] } \nonumber \\
{ {B}_{sig}^{q}(s,\Theta _{B}) } &=& { \frac{1}{24} [1+\cos ^{2}\Theta _{B}-\frac{4m^{2}}{%
s}\sin ^{2}\Theta _{B}] }
\end{eqnarray}
and for gluon-production
\begin{eqnarray}
{B}_{0}^{g}(s,\Theta _{B}) &=& \overline{c}(s,\Theta _{B})[\sin
^{2}\Theta _{B}(1+\cos
^{2}\Theta _{B})+\frac{8m^{2}}{s}(\cos ^{2}\Theta _{B}+\sin ^{4}\Theta _{B})-%
\frac{16m^{4}}{s^{2}}(1+\sin ^{4}\Theta _{B})] \nonumber \\
{B}_{sig}^{g}(s,\Theta _{B}) &=& \overline{c}(s,\Theta _{B})[\sin
^{2}\Theta _{B}(1+\cos
^{2}\Theta _{B})-\frac{8m^{2}}{s}( 1 +\sin ^{2}\Theta _{B})+%
\frac{16m^{4}}{s^{2}}(1+\sin ^{4}\Theta _{B})]
\end{eqnarray}
where the overall gluon-pole-factor
\begin{equation}
\overline{c}(s,\Theta _{B})=\frac{3 s^{2} g^4 }{96 (m^{2}-t)^{2}(m^{2}-u)^{2}} \,\, [7+%
\frac{36p^{2}}{s}\cos ^{2}\Theta _{B}]
\end{equation}
depends on the $(t \bar{t})_{c.m.}$ center-of-momentum energy
$\sqrt{s}$ and $\cos{\Theta_B}$, and includes the gluon color
factor. In application, for instance to $p p \rightarrow t \bar{t}
X$, parton-level top-quark spin-correlation functions need to be
smeared with the appropriate parton-distribution functions with
integrations over $\cos\Theta_B$ and the $ ( t \bar{t} )_{c.m.} $
energy, $\sqrt{s}$.

There is a common final-state interference structure in these
BR-S2SC functions for the charged-lepton plus jets reaction $p p $
or $p \bar{p} \rightarrow t \bar{t} \rightarrow \ldots $.  From
(11,12), the final-state relative phase effects do not depend on
whether the final $t_1 \overline{t}_2$ system has been produced by
gluon or by quark production.

Measurement of the sign of the $\eta_L \equiv \frac{\Gamma
_{R}(0,-1)}{\Gamma } = \pm 0.46$(SM/+) helicity parameter could
exclude a large chiral weak-transition moment in favor of the SM
prediction.

\section{ The three theoretical numerical puzzles }

These puzzles arose in a general search for empirical ambiguities
between the SM's $(V-A)$ coupling and possible single additional
Lorentz couplings that could occur in top-quark decay experiments,
see refs. in [1].

\bigskip

\textbf{1st Puzzle's associated phenomenological $m_{top}$ mass
formula: }

With ICHEP2008 empirical mass values ($m_{t} = 172.4\pm1.2GeV,
m_{W} = 80.413\pm0.048GeV$)
\begin{equation}
y=\frac{m_{W}}{m_{t}}=0.4664\pm 0.0035  \nonumber
\end{equation}

This can be compared with the amplitude equality in the upper part
of ``Table 1", see the corresponding two ``220" entries,
\begin{equation} A_{+} (0,-1/2) = a \, A_{SM} (-1,-1/2)  \nonumber
\end{equation}
with $a=1+O(v\neq y\sqrt{2},x)$. By expanding in the mass ratio
$x^{2}=(m_{b}/m_{t})^{2}=7\cdot 10^{-4}$,
\begin{eqnarray*}
1-\sqrt{2}y-y^{2}-\sqrt{2}y^{3}=x^{2}(\frac{2%
}{1-y^{2}}-\sqrt{2}y)-x^{4}(\frac{1-3y^2}{(1-y^2)^3})+\ldots \\
=1.89x^{2}-0.748x^{4}+\ldots \nonumber
\end{eqnarray*}
The only real-valued solution to this cubic equation is
$y=0.46006$ ($m_{b}=0$).

\bigskip

\textbf{Resolution of 2nd and 3rd Puzzles: }

In the lower part of ``Table 1", the two R-handed helicity b-quark
amplitudes $A_{New} = A_{g_L =1} / \sqrt \Gamma $ have the same
magnitude in the SM and the (+) model.

As a consequence of Lorentz-invariance, for the $t \rightarrow W^+
b$, the 4 intensity-ratios, $\, {\Gamma_{L,T}}|_{\lambda_b = \mp
\frac{1}{2}} / {\Gamma( t \rightarrow W^+ b )} \,$ are identical
for the standard model $V-A$ coupling and for the case of an
additional chiral weak-moment of relative strength $\Lambda_+ =
E_W / 2$.   This intensity-ratio equivalence does not depend on
the numerical values of $y=\frac{m_{W}}{m_{t}}$ or $x=
m_{b}/m_{t}$.

\section{ High energy properties of off-shell continuation }

A simple off-shell continuation of Eq(2) is the $t \rightarrow W^+
b$ vertex
\begin{equation}
\frac{1}{2} \Gamma ^\mu =g_L \left[ \gamma ^\mu P_L + \frac{m_t }
{ p_t \cdot q_W }\imath \sigma ^{\mu \nu } ({q_W})_\nu P_R \right]
\end{equation}
To lowest order, this additional coupling has good high energy
properties, i.e. it does not destroy 1-loop unitarity, for the
processes $t \overline{t} \rightarrow  W^{+}\ W^{-}$ and $b\
\overline{b}\rightarrow W^{-}\ W^{+}$ with both $W$-bosons with
longitudinal polarization, nor with either or both $W$-bosons with
transverse polarizations. There are of course no effects to lowest
 order for their analogues with a $Z^{o}Z^{o}$
final state, nor for\ the processes like $t\overline{\
b}\rightarrow e^{+}\nu _{e}$ . By power-counting, versus the SM's
fermion cancellations, there are no new effects for the ABJ gauge
anomalies.

The processes $t\overline{\ b}%
\rightarrow W^{+} \gamma, t\overline{\ b}
\rightarrow W^{+} H$ with $H$ a Higgs boson, are not divergent.
However, the additional gauge-vertex diagram for the process $t\overline{\ b}%
\rightarrow W^{+}\ Z^{o}$ with longitudinal gauge-boson
polarizations is linearly divergent in the large $E_{t}$\ limit in
the $(t\ \overline{b})_{cm}$ frame. \ Since this divergence
involves a denominator factor of  $ p_{t}\cdot (k_{W}+k_{Z})$,
instead of additional neutral current couplings to the
third-generation fermions, we introduce an additional more-massive
$X^{\pm }$ boson to cancel this divergence.  The additional $tXb$
vertex is
\begin{equation}
\frac{1}{2} \Gamma_X ^\mu = - g_L \left[ \frac{m_t } { p_t \cdot
q_X }\imath \sigma ^{\mu \nu } ({q_ X})_\nu P_R \right]
\end{equation}
and the additional $WXZ$ vertex is of the same structure as the
usual $WWZ$ vertex of the SM. This additional $tXb$ vertex does
not effect the SM's fermion cancellations of ABJ gauge anomalies.

\section{ Manifestation of intensity-ratio equivalence }

The $t \rightarrow W^+ b$ helicity decay amplitudes in the SM and
those with the additional (+) coupling of Eqs.(2, 18) are related
by a set of transformation matrices.  Similar to matrix
representations of Lie Groups, these matrices form various
algebras and sub-algebras, with associated symmetries which should
be useful in investigating deeper dynamics in top quark physics
[1,2,5].

This is an analytic generalization of the many numerical patterns
in Table 1, i.e. it isn't merely 2 isolated amplitudes being
related per Eq.(17). Instead, this makes manifest the
intensity-ratio equivalence of Sec. 7. The $tWb$-transformations
are: $A_{+}=v M $ $A_{SM}$, $v P $ $A_{SM}, \ldots $;
$A_{SM}=v^{-1} M $ $A_{+}$, $- v^{-1} P $ $A_{+}, \ldots $; $v$ is
the W-boson velocity in the t-quark rest frame; with $\Lambda_{+}
=E_W /2$ but unfixed values of $m_W/m_t$ and $m_b/m_t$.  Figs.
(4-5) give a compact and complete summary.
\begin{figure}[hbt]\centering
\begin{minipage}[c]{.45\linewidth}\centering
\includegraphics[width=7.95cm]{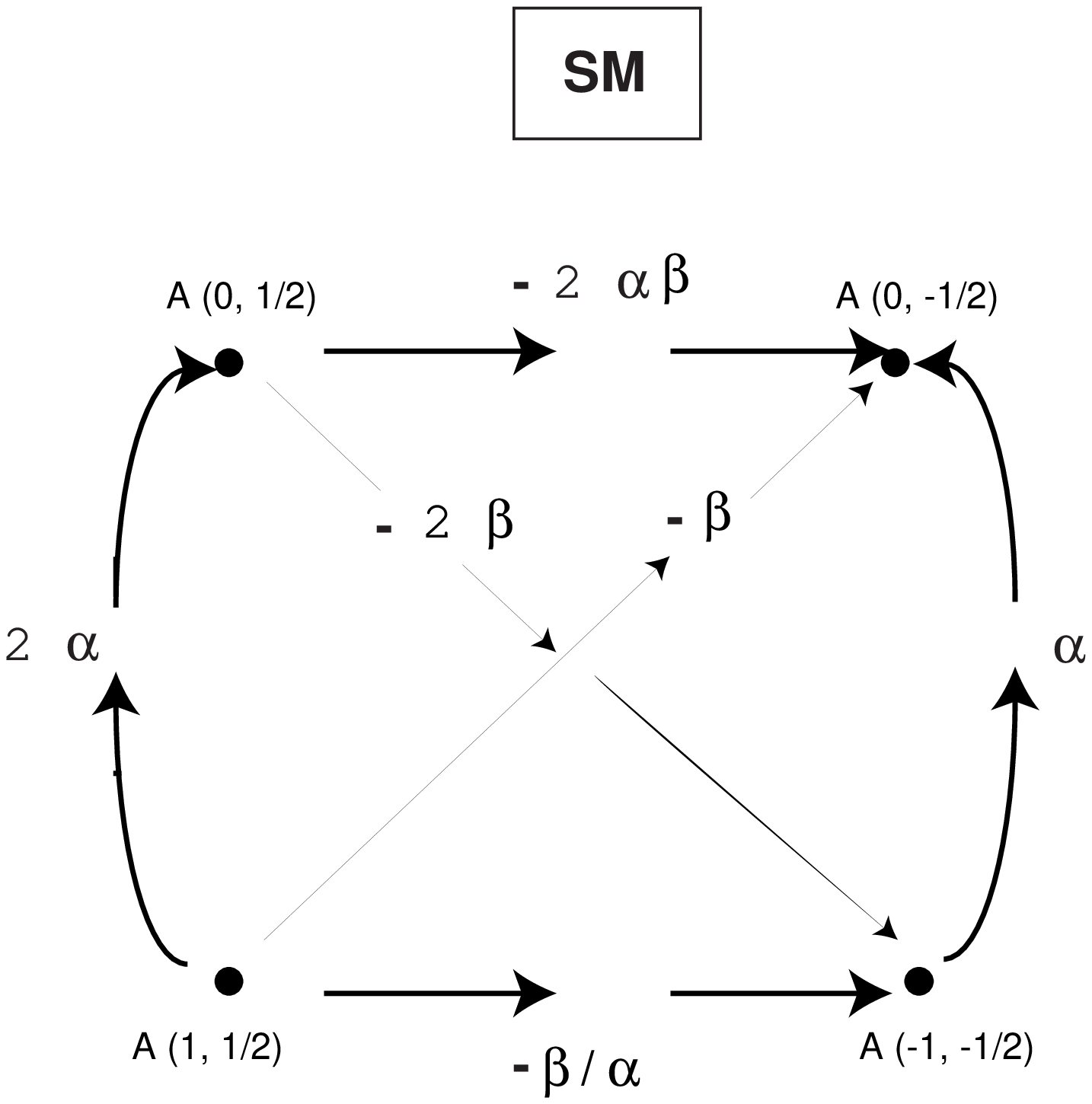}
\end{minipage}
\hskip 0.05cm
\begin{minipage}[c]{.45\linewidth}\centering
\includegraphics[width=7.95cm]{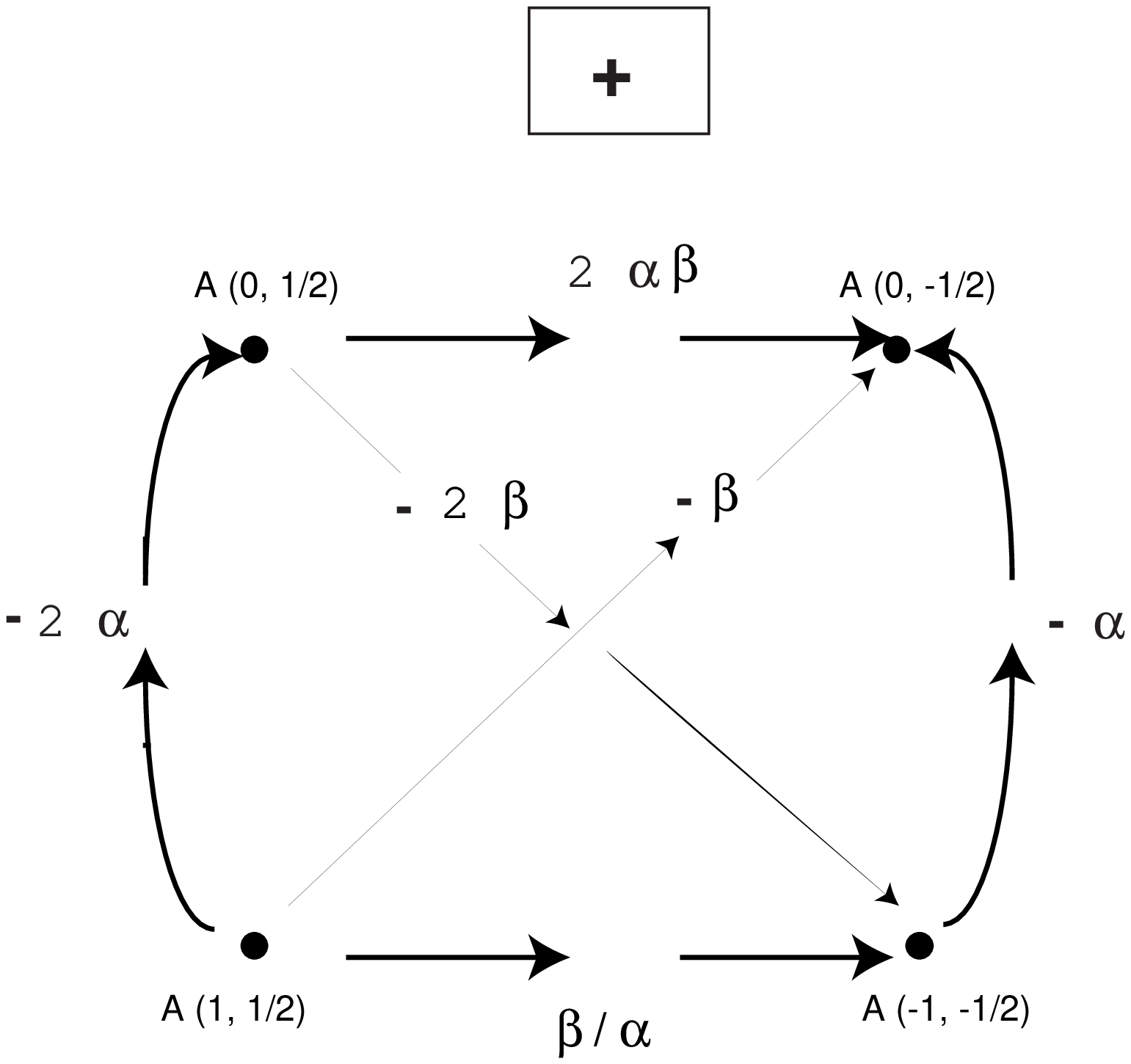}
\end{minipage}
\vskip -4.cm
 \caption{Left: Self-Transformation factors between the
helicity amplitudes of the Standard Model. Right:
Self-Transformation factors between the helicity amplitudes of the
(+) model; note the sign changes in the outer four factors versus
those for the SM.}
\end{figure}
\begin{figure}[hbt]\raggedright
\hskip -0.5cm
\includegraphics[width=14.cm]{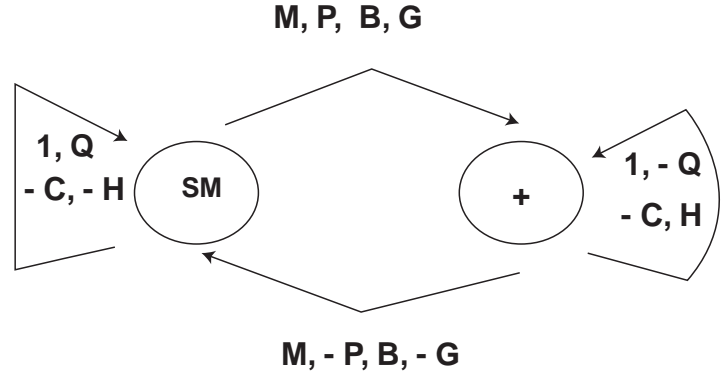}
\vskip 1.cm
 \caption{Diagram displaying the roles of all the
matrices in transforming the helicity amplitudes of either the SM
or the (+) model.  The matrices in the diagram form the 8-element
closed transformation algebra. Included are the sign factors
needed in order to get the correct transformation result using
each of the matrices.}
\end{figure}

With $\alpha = a/v, \beta = b/v$, these matrices are: $M=$
$diag(1,-1,-1,1)$,
\begin{equation}
P(\alpha) \equiv  \left[
\begin{array}{cccc}
0 & \alpha & 0 & 0 \\ - 1/\alpha & 0 & 0 & 0 \\ 0 & 0 & 0 & -1/2\alpha \\
0 & 0 & 2\alpha & 0
\end{array}
\right], \, B(\beta)\equiv  \left[
\begin{array}{cccc}
0 & 0 & 0 & -\beta \\ 0 & 0 & 2\beta & 0 \\ 0 & 1/2\beta & 0 & 0
\\ -1/\beta & 0 & 0 & 0
\end{array}
\right],
\end{equation}

\begin{equation}
G(\alpha,\beta) \equiv  \left[
\begin{array}{cccc}
0 & 0 & -2\alpha \beta & 0 \\ 0 & 0 & 0 & \beta / \alpha \\
1/2 \alpha \beta & 0 & 0 & 0
\\ 0 & -\alpha/\beta & 0 & 0
\end{array}
\right], \, Q(\alpha) \equiv  \left[
\begin{array}{cccc}
0 & \alpha & 0 & 0 \\ 1/\alpha & 0 & 0 & 0 \\ 0 & 0 & 0 & 1/2\alpha \\
0 & 0 & 2\alpha & 0
\end{array}
\right],
\end{equation}

\begin{equation}
C(\beta)\equiv  \left[
\begin{array}{cccc}
0 & 0 & 0 & \beta \\ 0 & 0 & 2\beta & 0 \\ 0 & 1/2\beta & 0 & 0
\\ 1/\beta & 0 & 0 & 0
\end{array}
\right], \, H(\alpha,\beta) \equiv  \left[
\begin{array}{cccc}
0 & 0 & 2\alpha \beta & 0 \\ 0 & 0 & 0 & \beta / \alpha \\
1/2 \alpha \beta & 0 & 0 & 0
\\ 0 & \alpha/\beta & 0 & 0
\end{array}
\right],
\end{equation}
Including the identity matrix, this is the closed 8-element
transformation algebra with a commutator/anticommutator structure.
It has $7=4_{-}+3_{+}$ three-element closed subalgebras ($\mp$
subscript denotes non-trivial commutators/anticommutators).  There
are 2x2 matrix compositions of the above matrices with associated
commutator/anticommutator algebras.

\begin{acknowledgments}
For their assistance in various stages of this research, we thank
E.G. Barbagiovanni, J.J. Berger, A. Borgia, P. Gan, R. Joachim,
N.H. Kim, S. Piotrowski, E.K. Pueschel, and J.R. Wickman.
\end{acknowledgments}

\end{document}